\documentclass[12pt]{article}
\usepackage[cp1251]{inputenc}
\usepackage[ukrainian]{babel}
\begin{document}
\large

\begin{center}
{\Large \bf Asymptotic soliton-like solutions to the singularly
perturbed Benjamin–Bona–Mahony equation with variable coefficients }
\end{center}

\begin{center}
\noindent Valerii Samoilenko, Yuliia Samoilenko\\
Department of Mathematical Physics, Taras Shevchenko National
University of Kyiv, Kyiv, Ukraine, 01601 \\
valsamyul@gmail.com, vsam@univ.kiev.ua, yusam@univ.kiev.ua
\end{center}

\vskip10mm

\begin{minipage}{150mm}
{\bf Abstract.} The paper deals with a problem of asymptotic
soliton-like solutions to the Benjamin–Bona–Mahony (BBM) equation
with a small parameter at the highest derivative and variable
coefficients depending on the variables $x $, $ t $ as well as a
small parameter. An algorithm for constructing solutions to the BBM equation
has been proposed and theorems on accuracy of such solutions have been proved.
\end{minipage}

\vskip10mm


\vskip5mm


\vskip10mm

{\bf 1. Introduction} In the modern mathematics and theoretical
physics, much attention is paid to the Korteweg-de Vries (KdV)
equation
\begin{equation} \label{KdV}
u_t + uu_x - u_{xxx} = 0
\end{equation}
as well as its different generalizations.

After the discovery of particular solutions to the KdV equation
describing waves with special features, which took place in 1965,
the equation has become an object for comprehensive study by many
researchers. These waves interact among themselves in a special way,
namely, they don't change their shapes after their collision. Such
particular solutions, called solitons \cite{Zabusky}, have been
studied mostly by physicists because soliton solutions describe
propagation of long waves in shallow water as well as in various
media. Later, similar soliton solutions have been also found for a
variety of nonlinear dynamical systems, in particular for the
modified KdV equation \cite{Wadati}, the nonlinear Shrodinger
equation \cite{Toda} and the derivative Shrodinger equation
\cite{KaupN}, the Kadomtsev-Petviashvili equation \cite{KadPetv},
the Kaup system \cite{Kaup}, and many others. At present, the
soliton theory is widely applied in hydrodynamics, plasma physics,
nonlinear optic, quantum field theory, solid physics, biology, etc.

A number of monographs and numerous papers are devoted to
different aspects of the soliton theory. Numerous works investigated
different analytical and qualitative properties
of solutions to the KdV equation and its diverse generalizations using various methods and
approaches \cite{Zabusky}, \cite{GGKM} -- \cite{Blacmore}, including
numerical methods, that are usually used for researching complex
nonlinear models.

Studies of the KdV equation stimulated efforts to find new
nonlinear systems possessing similar properties, namely the propagation of long waves and the
existence of soliton solutions.
In 1966, Peregrine~D.H. \cite{Peregrin} proposed the following equation
\begin{equation}\label{BBM_0}
u_t + u_x + uu_x - u_{xxt} = 0
\end{equation}
as an alternative to the KdV equation.
Equation (\ref{BBM_0}) describes the propagation of long waves in
nonlinear dispersive media and has soliton solutions. It was
thoroughly  studied by Benjamin~T.B., Bona~J.L., and Mahony~J.J. in
\cite{BBM}. These researchers have demonstrated the existence of
classical solutions to (\ref{BBM_0}) and the uniqueness of solution
to the initial-value problem for it. In \cite{BBM}, it was also
proved that solutions depend continuously on the initial values as well as
on the forcing functions added to the right-hand side of equation
(\ref{BBM_0}). In other words, the initial-value problem for
equation (\ref{BBM_0}) is confirmed to be classically well possed in
the Hadamard sense.

Equation (\ref{BBM_0}) was originally called the regularized long wave equation
\cite{BBM}. At present, equation (\ref{BBM_0}) is also called the
Benjamin–Bona–Mahony equation or the BBM equation.

After the initial publication \cite{BBM}, equation (\ref{BBM_0}) has been
discovered to describe a variety of physical phenomena and processes, in
particular, the propagation of long waves in liquids, acoustic waves in
anharmonic crystals, acoustic-gravity waves in compressible fluids
and others \cite{Seadway}. Subsequent studies showed that equation
(\ref{BBM_0}) possesses numerous interesting
properties. It was stated in \cite{Eilback1, Eilback2} that equation (\ref{BBM_0}) has, in some
sense, "much nicer mathematical properties than the KdV equation".

Indeed, the KdV equation and the BBM equation have several different
mathematical properties. First of all, it should be noted the BBM
equation is more convenient for studying through numerical technique
than the KdV equation because the numerical schemes \cite{Peregrin},
\cite{Eilback1, Eilback2} applied for (\ref{BBM_0}) are stable for
greater time steps \cite{Dodd} compared to numerical schemes
proposed for (\ref{KdV}). That is why the BBM equation was first
intensively studied through different numerical methods. As a
result, many improbable properties of equation (\ref{BBM_0}) have
been found. For example, in the case of the BBM equation, the
numerical study  of the two-soliton collision and three-soliton
collision appointed "collisional stability"\, or elastic interaction
\cite{Eilback1, Eilback2}. In other words, it was numerically
demonstrated that after the nonlinear interaction, the solitary
waves again have their initial amplitudes.

On the other hand, the inelastic collision of
two solitary waves moving in opposite directions was also found \cite{Santarelli}.
Moreover, the BBM equation has only three conservation laws while
the KdV equation has an infinite conservation law hierarchy. As a
consequence, the equation (\ref{BBM_0}) is not integrable.

Because of the numerical results on the existence of both the two- and
three-soliton solutions to the BBM equation, it is natural to
clarify whether the equation has an analytical $N$-soliton solution
as the KdV equation. Many authors studied the problem through
various techniques but without success \cite{Santarelli} --
\cite{Bona2}. At present it is known that the BBM equation has
neither exact two-soliton solution nor exact multi-soliton solution
[12, p.~649].

At the same time, a number of exact and numerical solutions and new
qualitative properties to the BBM equation have been discovered. For
example, different exact solutions to (\ref{BBM_0}) were found in
\cite{Seadway,Arora, Wazwaz}  using the {\it sech -- tanh}-method
and the {\it cos}-function method. In particular, it was found
through the {\it sech -- tanh}-method that the BBM equation has a
soliton of the following form \cite{Wazwaz}
\begin{equation}\label{one-phase}
u(x, t) = 3(a-1) \, ch^{-2} \left(\frac{1}{2} \sqrt{\frac{a-1}{a}}
(x - at) + C \right),
\end{equation}
where $ a $, $ C $ are some real constants. %  and $ | a | > 1 $.

Moreover, shock solutions to equation (\ref{BBM_0}) where found in
\cite{El}, asymptotic stability of its solitary waves was studied in
\cite{Mizumachi}, the existence of the global attractor for it was
proved in \cite{Wang}, and in \cite{Avrin}, the existence of the
global solutions to the BBM equation was established.

Taking into account that equation (\ref{BBM_0}) has one-soliton
solution mentioned above, it's naturally to consider the problem of
constructing asymptotic soliton-like solutions to the BBM equation
with variable coefficients and a small parameter at the highest
derivative analogously to the KdV equation \cite{Sam1} --
\cite{Sam3}.

Thus, the present paper deals with the singularly perturbed equation
of the following form
\begin{equation}\label{BBM_v}
a(x, t, \varepsilon) u_t + b(x, t, \varepsilon) u_x + c(x, t,
\varepsilon) uu_x - \varepsilon^n u_{xxt} = 0, \quad (x, t) \in
{\mathbf R}\times [0; T],
\end{equation}
where  $ \varepsilon > 0 $ is a small parameter, $ n $ is natural
and functions $ a(x, t, \varepsilon) $, $ b(x, t, \varepsilon ) $, $
c(x, t, \varepsilon ) $ are generally infinitely differentiable with
respect to all variables $ (x, t, \varepsilon) \in {\mathbf R}
\times [0; T] \times [0; \varepsilon_0) $ for some $T>0$,
$\varepsilon_0>0$.

The functions $ a(x, t, \varepsilon) $, $ b(x, t, \varepsilon ) $, $
c(x, t, \varepsilon ) $ are supposed to have asymptotic expansions
$$
a(x, t, \varepsilon) = \sum\limits_{k=0}^N \varepsilon^k a_k(x, t) +
O(\varepsilon^{N+1}) ,
$$
$$
b(x, t, \varepsilon) = \sum\limits_{k=0}^N \varepsilon^k b_k(x, t) +
O(\varepsilon^{N+1}),
$$
\begin{equation} \label{coeff}
c(x, t, \varepsilon) = \sum\limits_{k=0}^N \varepsilon^k c_k(x, t) +
O(\varepsilon^{N+1}) .
\end{equation}
In addition we suppose $ a_0(x, t) \, b_0(x, t) \, c_0(x, t) \not= 0
$ for all $ (x, t) \in {\mathbf R} \times [0; T] $.

We search functions represented as asymptotic expansions in a small
parameter that satisfy equation (\ref{BBM_v}) with certain
accuracy. These functions are chosen so that in the case of constant
coefficients, they coincide with the exact soliton solutions of
equation (\ref{BBM_v}). Therefore, the searched functions are called
asymptotic soliton-like solutions to the given equation.

To find such functions, general ideas and methods
of asymptotic analysis can be applied. We need to define the form of these
functions, to propose an recurrent algorithm  determining
all of the members of the corresponding asymptotic solution,  and, in
addition, to evaluate the accuracy with which the asymptotic
approximations satisfy the equation.

The problem of the existence of solutions to a equation is not
usually studied in asymptotic analysis \cite{Ablowitz2},
\cite{Naife}, \cite{Maslov2}, \cite{Maslov3}, \cite{Maslov1},
\cite{Glebov}, \cite{Kalyakin}, since this question relates to
problems of another type, which are usually complex and require
completely different methods and approaches for their analysis.

For example, for studying the problem of existence of solutions to the
generalized KdV equation with variable coefficients in the Schwartz
space, methods of parabolic regularization and a priori
estimates \cite{Fam} are used.

Below, we present an algorithm for constructing asymptotic
soliton-like solutions to equation (\ref{BBM_v}) and
find the accuracy with which constructed asymptotic solutions
satisfy equation (\ref{BBM_v}). The algorithm is analogous to the
one developed for constructing asymptotic one-, two- and multi-phase
soliton-like solutions to the KdV equation with variable
coefficients \cite{Sam1, Sam2, Sam3}. It is based on the nonlinear WKB
technique. It should be noted that asymptotic
soliton-like solutions to the integrable type equations were firstly
constructed by Maslov~V.P. and his coauthors \cite{Maslov1}. They
applied the nonlinear WKB technique developed for constructing
quasi-periodic solutions to the singularly perturbed KdV equation
with constant coefficient \cite{Miura} in 1974.

The present paper is organized as follows. In Section 2, we give
preliminary notes and formulate auxiliary definitions. In Section 3,
an algorithm for constructing an asymptotic soliton-like
solution to the BBM equation is proposed and described in
details. We discuss procedures of finding terms of asymptotic
expansions and, in particular, solvability of differential equations
for regular and singular parts of the asymptotic solutions. In
Section 4, theorems on justification of proposed
algorithm are proved.

{\bf 2. Preliminary notes and definitions}

Let $ S = S({\mathbf R}) $ be a space of quickly decreasing
functions, i.e. the space of infinitely differentiable on $ {\mathbf
R} $ functions such that for any integers $ m, n \ge 0 $ the
condition
$$
\sup\limits_{x \in {\mathbf R} } \left| \, x^m \, D^n u (x) \,
\right| < + \infty
$$
is satisfied.

Let $ G_1 = G_1({\bf R} \times [0; T] \times {\bf R}) $ be a space
of infinitely differentiable functions $ f = f(x, t, \tau) $, $ (x,
t, \tau) \in {\bf R} \times [0; T] \times {\bf R} $ such that there
are fulfilled the following conditions \cite{Maslov1}:

$1^0$. the relation
$$ \lim\limits_{\tau \to + \infty} \tau^n
\frac{\partial\,^p}{\partial x^p} \, \frac{\partial \, ^q}{\partial
\, t^q} \, \frac{\partial \,^r}{\partial \tau^r} \, f (x, t, \tau) =
0, \quad (x, t) \in K,
$$
takes place;

$2^0$. there exists such a differentiable function $ f^-(x, t) $
that on any compact set $ K \subset {\bf R} \times [0; T] $
condition
$$
\lim\limits_{\tau \to - \infty} \tau^{n} \frac{\partial
\,^p}{\partial \, x^p} \, \frac{\partial \, ^{q}}{\partial \, t^{q}}
\, \frac{\partial \,^{r}}{\partial \, \tau^{r}} \, \left( f (x, t,
\tau) - f^{-}(x, t)\right) = 0, \quad (x, t) \in K,
$$
is satisfied for any non-negative integers $ n $, $ p $, $ q $, $ r
$ uniformly with respect to $ (x, t) \in K $.

Let $ G_1^0 = G_1^0 ({\bf R} \times [0; T] \times {\bf R}) \subset
G_1 $ be a space of functions $ f = f(x, t, \tau) \in G_1 $, $ (x,
t, \tau) \in {\bf R}\times [0; T] \times {\bf R} $ such that
uniformly with respect to variables $ (x, t) $ on any compact $ K
\subset {\bf R} \times [0; T] $ the following condition
$$
\lim_{\tau \to \, - \infty} f (x, t, \tau) = 0
$$
takes place.

We use the following definition of an asymptotic soliton-like
function.

{\bf Definition}. {\it A function  $ u = u(x, t, \varepsilon) $,
where $ \varepsilon $ is a small parameter, is called an asymptotic
one phase soliton-like function if for any integer $ N \ge 0 $ it
can be represented in the form of
\begin{equation}\label{2as_sol}
u(x, t, \varepsilon) = \sum\limits_{j=0}^N \varepsilon^j
\left[u_j(x, t) + V_j (x, t, \tau)\right] + O(\varepsilon^{N+1}),
\quad \tau = \frac{x - \varphi(t)}{\varepsilon},
\end{equation}
where $ \varphi(t) \in C^{\infty} ([0;T]) $ is a scalar real-valued
function; \, $ u_j(x, t) \in C^\infty ({\mathbf R}\times [0; T]) $,
$ j = \overline{0, N }; $ \, $ V_0 (x, t, \tau) \in G_1^0; $ \, $
V_j(x, t, \tau) \in G_1 $, $ j = \overline{1\,, N }$.}

The function $ x - \varphi(t) $ is called a phase of the one-phase
soliton-like function $ u(x, t, \varepsilon). $ A curve determined
by equation $ x - \varphi(t) = 0 $ is called a curve of
discontinuity for function (\ref{2as_sol}).

It should be noted that the definition 1 was formulated in monograph
\cite{Maslov1} concerning problem of constructing asymptotic
soliton-like solution to the KdV equation with small parameter at
the highest derivative for the case when power of small parameter is
equal to 2. In general, an asymptotic soliton-like solution to the
singularly perturbed KdV equation has more complicated structure
\cite{Sam1} depending on the power of a small parameter at the
highest derivative.

In the sequel, we use the notation of asymptotic analysis of the
following form $ \Psi (x,t, \varepsilon ) = O\left( \varepsilon^N
\right) $, $ \varepsilon \to 0$. It means that there exist such
values $ \varepsilon_0 > 0 $, $ C > 0 $ that $ | \Psi (x,t,
\varepsilon ) | \le C \, \varepsilon^N $ for all $ \varepsilon \in
(0; \varepsilon_0)$, $ (x,t) \in K $, where compact set $ K \subset
{\bf R} \times [0;T] $ and value $ C $ is only depending on the
number $ N $ and the set $ K \subset {\bf R} \times [0;T] $.

{\bf 3. Algorithm for constructing asymptotic solutions.} Let us
consider a problem of constructing asymptotic one-phase soliton-like
solutions to equation (\ref{BBM_v}). Analogously to the case of the
KdV equation \cite{Sam1, Sam2, Sam3}, the form of the asymptotic
solutions depends on the power $ n $ of a small parameter at the
highest derivative in (\ref{BBM_v}). So, asymptotic solutions to
equation (\ref{BBM_v}) are written in the form
\begin{equation} \label{sol_one-phase}
u(x, t, \varepsilon) = \sum\limits_{j=0}^N \varepsilon^j u_j(x, t) +
\sum\limits_{j=0}^N \varepsilon^j V_j (x, t, \tau) +
O(\varepsilon^{N+1}), \quad \tau =
\frac{x-\varphi(t)}{\varepsilon^{{n}/{2}}},
\end{equation}
when $ n $ is an even number, and
$$
u(x,t, \varepsilon)= \sum\limits_{j=0}^{k} \varepsilon^j u_j(x,t) +
\sum\limits_{j=0}^{k} \varepsilon^j V_j (x, t, \tau) + \varepsilon^k
\sum\limits_{j=1}^{2N-2k} \varepsilon^{j/2} u_j(x,t) +
$$
\begin{equation} \label{sol_one-phase_odd}
+ \varepsilon^k \sum\limits_{j=1}^{2N-2k} \varepsilon^{j/2} V_j (x,
t, \tau) + O(\varepsilon^{N+\frac{1}{2}}), \quad \tau =
\frac{x-\varphi(t)}{\sqrt{\varepsilon} \,\, \varepsilon^{k}},
\end{equation}
when $ n $ is such an odd number that $ n=2k+1$, $k \in \mathbf{N}
\cup \{ 0 \}$.

Further, we consider the case $ n = 2 $ allowing us to demonstrate
all details of algorithm on searching asymptotic solutions to
equation (\ref{BBM_v}). Thus, the asymptotic solutions are
constructed as follows
\begin{equation} \label{sol_one-phase}
u(x,t,\varepsilon) = Y_N(x,t,\tau, \varepsilon) +
O(\varepsilon^{N+1}),
\end{equation}
where
$$
Y_N (x, t, \tau, \varepsilon) = \sum\limits_{j=0}^N \varepsilon^j
\left[ u_j(x, t) + V_j(x, t, \tau) \right], \quad \tau =
\frac{x-\varphi(t)}{\varepsilon}.
$$
The function
\begin{equation}  \label{regular_part}
U_N (x, t, \varepsilon) =  \sum\limits_{j=0}^N \varepsilon^j u_j(x,
t)
\end{equation}
is called a regular part of asymptotic solution
(\ref{sol_one-phase}), and the function
\begin{equation} \label{singular_part}
V_N (x, t, \tau, \varepsilon) = \sum\limits_{j=0}^N \varepsilon^j
V_j(x, t, \tau)
\end{equation}
is called a singular part of asymptotic solution
(\ref{sol_one-phase}).

It's clear $ Y_N (x, t, \tau, \varepsilon) = U_N (x, t, \varepsilon)
+ V_N(x, t, \tau, \varepsilon). $

In order to obtain differential equations for terms of the regular
and the singular parts of asymptotic series (\ref{sol_one-phase}), we
apply basic ideas of asymptotic analysis \cite{Naife},
\cite{deBrain}, \cite{Bogoliu_Mitropol}. In more details, we put
series (\ref{sol_one-phase}) into equation (\ref{BBM_v}), take into
account property $ V_j(x, t, \tau) \in G_1 $, $j=\overline{0,N}$,
and equate coefficients at the same powers of a small parameter in
left and right sides of the relation after the substitution.

The terms of the regular part (\ref{regular_part}) satisfy the following
system of the first order partial differential equations
\begin{equation} \label{reg_part_0}
a_0(x, t) \frac{\partial u_0}{\partial t} +
b_0(x, t) \frac{\partial u_0}{\partial x} + c_0(x, t) u_0
\frac{\partial u_0}{\partial x} = 0,
\end{equation}
$$
a_0(x, t) \frac{\partial u_j}{\partial t} + b_0(x, t) \frac{\partial
u_j}{\partial x} + c_0(x, t) \left(u_j \frac{\partial u_0}{\partial
x}  + u_0 \frac{\partial u_j}{\partial x} \right) =
$$
\begin{equation} \label{reg_part_1}
= f_j(x, t, u_0, u_1, \ldots , u_{j-1}), \quad j = \overline{1, N},
\end{equation}
where functions $ f_j(x, t, u_0, u_1, \ldots , u_{j-1}), $ $ j =
\overline{1, N} $, are recurrently determined.

System (\ref{reg_part_0}), (\ref{reg_part_1}) contains
the only quasi-linear equation while the others are linear ones. The
functions $ u_j(x, t), $  $ j = \overline{0, N} $, can be easy
found recurrently through integrating equations (\ref{reg_part_0}),
(\ref{reg_part_1}), for example, by means of the method of
characteristics \cite{Evans}. Regarding it the terms of regular part
(\ref{regular_part}) of asymptotic are supposed to be known.

\textbf{3.1. Singular part of asymptotic.} The terms of singular
part (\ref{singular_part}) are defined as solutions to system of the
third order partial differential equations
\begin{equation} \label{singular_part_0}
\varphi' (t) \, \frac{\partial^3 V_0}{\partial \tau^3} + \left(
b_0(x, t) - \varphi' (t) \, a_0(x, t) \right) \, \frac{\partial V_0}
{\partial\tau} + c_0(x, t) \left( u_0  + V_0 \right)\frac{\partial
V_0}{\partial\tau} = 0,
\end{equation}
$$
\varphi' (t) \, \frac{\partial^3 V_j}{\partial \tau^3} + \left(
b_0(x, t) - \varphi' (t) \, a_0(x, t) \right) \frac{\partial V_j}
{\partial\tau} + c_0(x, t) \left( u_0 \frac{\partial V_j}
{\partial\tau} + \frac{\partial}{\partial \tau} \left( V_0 V_j
\right) \right) =
$$
\begin{equation} \label{singular_part_1}
= {F}_j (x, t, \tau),
\end{equation}
where functions $ {F}_j (x,t, \tau)=F_j(t, V_0(x, t, \tau), \ldots ,
V_{j-1} (x, t, \tau)$, $u_0(x, t)$, $\ldots $, $u_j(x, t)), $ are
defined recurrently after determining the functions $ u_0(x,t) $, $
u_1(x,t) $, $\ldots $, $ u_j(x,t)$, $ V_0(x,t, \tau), $ $ V_1(x, t,
\tau) $, $ \ldots $, $ V_{j-1} (x,t, \tau)$, $j= \overline{1,N}$.

Recall that the solutions to equations (\ref{singular_part_0}),
(\ref{singular_part_1}) must belong to the spaces $G_1^0$, $G_1$
correspondingly. Besides, under the searching the functions $ V_{j}
(x,t, \tau)$, $j= \overline{0, N}$, we have also to find a function
$ \varphi = \varphi(t) $ defining a discontinuity curve $ \Gamma =
\{ (x,t) \in \mathbf{R} \times [0;T]: \, x = \varphi(t) \} $.

Taking the remarks into attention, we may study system
(\ref{singular_part_0}), (\ref{singular_part_1}) as follows. Firstly,
we assume the function $ \varphi = \varphi(t) $ is known. Then
equations (\ref{singular_part_0}), (\ref{singular_part_1}) are
considered on the discontinuity curve $ \Gamma $ and value $ t $ in
the equations is sup\-po\-sed to be a parameter. In this connection
the function $ v_0(t, \tau) = V_0 (x, t, \tau) \biggr|_{ \, x =
\varphi(t)} $ can be found in explicit form.

Secondly, we prove $ v_0(t, \tau) $ to be a quickly decreasing
function with respect to variable $ \tau $, i.e. $ v_0(t, \tau) \in
G_1^0 $. Then using property $ V_1(x, t, \tau) \in G_1 $, we find the
solution $ v_1(t, \tau) = V_1 (x, t, \tau) \biggr|_{ \, x =
\varphi(t)} $ in explicit form too. Moreover, we receive necessary
and sufficient condition on existence the solution in the space of
quickly decreasing functions with respect to the variable $ \tau $
as $ \tau \to +\infty $. Later, the condition is used for obtaining
nonlinear ordinary differential equation for function $ \varphi =
\varphi(t) $.

We now proceed with the description of the algorithm in details. Denote $ v_j
= v_j(t, \tau) = V_j (x, t, \tau) \biggr|_{\, x = \varphi(t)}, $ $ j
= \overline{0, N} $. From (\ref{BBM_v}), (\ref{singular_part_0}) it
follows that functions $ v_j $, $ j = \overline{0, N} $, satisfy
differential equations:
\begin{equation} \label{sing_part_02}
\varphi'(t) \, \frac{\partial^3 v_0}{\partial \tau^3} +
\left(b_0(\varphi, t) - \varphi' (t) \, a_0(\varphi, t) +
c_0(\varphi, t) u_0(\varphi, t)\right) \frac{\partial
v_0}{\partial\tau} + c_0(\varphi, t) \, v_0 \frac{\partial
v_0}{\partial\tau}  = 0;
\end{equation}
$$
\varphi' (t) \, \frac{\partial^3 v_j}{\partial \tau^3} + \left(
b_0(\varphi, t) - \varphi' (t) \, a_0(\varphi, t)  + c_0(\varphi, t)
u_0(\varphi, t) \right) \, \frac{\partial v_j}{\partial\tau} +
$$
\begin{equation}\label{sing_part_j2}
+ c_0(\varphi, t) \frac{\partial }{\partial\tau} \left(v_0
v_j\right) = {\cal F}_j (t, \tau),
\end{equation}
where $ {\cal F}_j (t, \tau) = F_j(t, V_0(x, t, \tau), \ldots ,
V_{j-1} (x, t, \tau), u_0(x, t), \ldots , u_j(x, t)) \biggr|_{ \, x
= \varphi(t)}, $ are easy defined recurrently after finding
functions $ u_0(x,t) $, $ u_1(x,t) $, $\ldots $, $ u_j(x,t)$, $
V_0(x,t, \tau), $ $ V_1(x, t, \tau) $, $ \ldots $, $ V_{j-1} (x,t,
\tau)$, $j= \overline{1,N}$. Here and below $ \varphi = \varphi(t)
$.

In (\ref{sing_part_j2}), for example,
$$
{\cal F}_1 (t, \tau) = - a_0(\varphi, t) \frac{\partial
v_0}{\partial t} - c_0(\varphi, t) u_{0x}(\varphi, t) \, v_0 -
$$
$$
- \left(  c_{0x}(\varphi, t) u_0 (\varphi, t) + c_0(\varphi, t)
u_{0x} (\varphi, t) - \varphi' (t) \, a_{0 x}(\varphi, t)  + b_{0
x}(\varphi, t) \right) \tau \frac{\partial v_0}{\partial \tau} -
$$
\begin{equation}\label{function_F}
- \left(c_{0 x}(\varphi, t) \tau + c_{1}(\varphi, t) \right) v_0
\frac{\partial v_0}{\partial \tau} -
\end{equation}
$$
- \left(c_0(\varphi, t) u_1 (\varphi, t) + c_{1}(\varphi, t) u_0
(\varphi, t) - \varphi' (t) \, a_1(\varphi, t) + b_1(\varphi,
t)\right) \frac{\partial v_0}{\partial\tau} + \frac{\partial^3
v_0}{\partial\tau^2 \partial t}.
$$

Let us consider equation (\ref{sing_part_02}). We can find its
solution in the following way. By integrating it with respect to $
\tau$, we obtain
\begin{equation}\label{sing_part_0_12}
\varphi' (t) \, \frac{d^2 v_0}{d\tau^2} = A(\varphi, t) v_0(t, \tau)
- \frac{1}{2} \, c_0(\varphi, t) v_0^2 (t, \tau) + C_1(t),
\end{equation}
where
\begin{equation} \label{A}
A(\varphi, t) = \varphi' (t) \, a_0(\varphi, t) - b_0(\varphi, t) -
c_0(\varphi, t) u_0 (\varphi, t).
\end{equation}
Since $ v_0(t, \tau) \in G_1^0 $ we can put $ C_1 (t)\equiv 0. $

Multiplying both sides of equation (\ref{sing_part_0_12}) by $ {d
v_0}/{d \tau} $ gives us
$$
\frac{1}{2} \, \varphi' (t) \, \frac{d}{d \tau} \left( % \left(
\frac{d v_0}{d\tau}\right)^2 % \right)
= \frac{1}{2} \, A (\varphi, t) \frac{d v_0^2}{d \tau} - \frac{1}{6}
\, c_0(\varphi, t) \frac{d v_0^3}{d \tau}.
$$

By further integration of the relation in $ \tau $ we get
\begin{equation} \label{sing_part_0_22}
\frac{1}{2} \, \varphi' (t) \, \left( \frac{d v_0}{d\tau}\right)^2 =
\frac{1}{2} \, A(\varphi, t) v_0^2 - \frac{1}{6} \, c_0(\varphi, t)
v_0^3 + C_2(t).
\end{equation}

Property $ v_0(t, \tau) \in G_1^0 $ allows us to take $ C_2(t)
\equiv 0.$

Finally, we have solution to equation (\ref{sing_part_02}) in the
following form
\begin{equation} \label{2sol_sing}
v_0(t, \tau ) = \frac{3 A(\varphi,t)}{c_0(\varphi,t)} \, ch^{-2}
\left( \frac{1}{2} \, \, \sqrt{\frac{A(\varphi, t)}{\varphi'
(t)}}(\tau + C_0(t)) \right) .
\end{equation}

Function (\ref{2sol_sing}) is evidently quickly decreasing with
respect to variable $ \tau $ for all $ t \in [0; T] $.

Thus the next lemma is true.

{\bf Lemma 1.} {\it If inequality
\begin{equation} \label{A>0}
\varphi'(t) A(\varphi(t), t) > 0
\end{equation}
is fulfilled for all $ t \in [0; T] $, then function
(\ref{2sol_sing}) is a solution to equation (\ref{sing_part_02}) and
belongs to the space $ G_1^0 $.}

\textbf{3.2. Orthogonality condition.} Let us discuss now the problem
on existence of solution to equation (\ref{sing_part_j2}) in the
space $ G_1 $.

The following lemma ensures necessary and sufficient conditions of
solvability of the problem.

{\bf Lemma 2.} {\it  Let inequality (\ref{A>0}) be fulfilled for all
$ t \in [0; T] $ and $ {\cal F}_j(t, \tau) \in G_1^{\, 0} $, $ j =
\overline{1,N} $. Then equation (\ref{sing_part_j2}) has a solution
$ v_j(t, \tau) \in G_1 $, $ j =\overline{1, N} $, if and only if
\begin{equation}\label{ort_cond}
\int\limits_{-\infty}^{+\infty} {\cal F}_j(t, \tau) v_0(t, \tau) d
\tau = 0, \quad j =\overline{1, N}.
\end{equation}
}

{\bf Proof.} Firstly, we show that a solution to equation
(\ref{sing_part_j2}) can be represented as follows
\begin{equation}\label{vyglyad}
v_j(t, \tau) = \nu_j(t) \eta_j(t, \tau) + \psi_j(t, \tau), \quad j =
\overline{1,N} ,
\end{equation}
where $ \eta_j(t, \tau) \in G_1 $ \, and \, $ \lim\limits_{\tau\to
-\infty} \eta_j(t, \tau) = 1 $; \, $ \psi_j(t, \tau) \in G_1^0 $;
\begin{equation}\label{nu}
\nu_j(t) = [- \varphi' a_0(\varphi, t) + b_0(\varphi, t) +
c_0(\varphi, t) u_{0}(\varphi, t)]^{-1} \lim\limits_{\tau\to-\infty}
\Phi_j(t, \tau),
\end{equation}
\begin{equation}\label{Phi}
\Phi_j(t, \tau) = \int\limits_{-\infty}^{\tau} {\cal F}_j(t, \xi)
d\xi + E_j(t).
\end{equation}
Here value $ E_j(t) $ no depends on variable $ \tau $ and it can be
found from condition
$$
\lim\limits_{\tau\to +\infty}\Phi_j(t, \tau) = 0.
$$

To state representation (\ref{vyglyad}) we integrate equation
(\ref{sing_part_j2}) in $ \tau$ in limits from $-\infty$ to $\tau $.
So, we obtain differential equation
\begin{equation}\label{ad_eq_1}
L v_j = \Phi_j(t, \tau),
\end{equation}
where operator $ L $ is defined with formula
\begin{equation}\label{operatorL_1}
L = \varphi' (t) \, \frac{d^2}{d\tau^2} \, - \, \varphi' (t) \,
a_0(\varphi, t) + b_0(\varphi, t) + c_0(\varphi, t) u_0(\varphi, t)
+ c_0(\varphi, t)v_0(t, \tau).
\end{equation}

By virtue of formulae (\ref{vyglyad}), (\ref{ad_eq_1}) the function
$ \psi_j (t, \tau ) $, $j=\overline{1,N},$ has to satisfy
inhomogeneous equation
\begin{equation}
\label{ad_eq_2} L \psi_j = \Phi_j - \nu_j L \eta .
\end{equation}

Owing to property $ {ker}~L^* = \{v_{0\tau}\} $ and theorem on
existence of a solution to the inhomogeneous equation with the
one-dimensional Schrodinger operator in the space of quickly
decreasing functions \cite{Sam4}, we obtain the following statement:
equation (\ref{ad_eq_2}) has a solution in the space $ G_1^0 $ iff
the following orthogonality condition
\begin{equation} \label{orth_cond_1}
\int\limits_{-\infty}^{+\infty} \left( \Phi_j - \nu_j L \eta \right)
v_{0\tau} d \tau = 0, \quad j=\overline{1,N},
\end{equation}
takes place.

Finally, from (\ref{orth_cond_1}), (\ref{Phi}), (\ref{ad_eq_1}), we
deduce condition (\ref{ort_cond}).

Lemma 2 is proved.

Relation (\ref{ort_cond}) is called orthogonality condition. In case
$ j =1 $ it may be used for deducing differential equation for
function $ \varphi = \varphi(t) $. The condition can be also used
for determining an interval $ [0; T] $ where asymptotic one-phase
soliton-like solution has to be considered. The problem is studied
in details below.

The following lemma describes more exact properties of solution to
equation (\ref{sing_part_j2}).

{\bf Lemma 3.} {\it Let conditions of lemma 2 and relation
(\ref{ort_cond}) be satisfied. Then $ v_j(t, \tau) \in G_1^0$,
$j=\overline{1,N}, $ if and only if the condition
\begin{equation} \label{ort_cond_2}
\lim\limits_{\tau \to -\infty} \Phi_j(t, \tau ) = 0, \quad j =
\overline{1,N},
\end{equation}
is fulfilled.}

The proof is trivial and it follows from representation
(\ref{vyglyad}).

In particular case as $ j=1 $ condition (\ref{ort_cond_2}) has the
following form
$$
a_0 (\varphi, t)\frac{d}{d t} \frac{\sqrt{A(\varphi, t) \varphi{\,'}
(t)}}{c_0(\varphi, t)} + \frac{\sqrt{A(\varphi, t) \varphi{\,'} (t)
}}{c_0(\varphi, t)} \times
$$
\begin{equation} \label{cond_G_0}
\times \left[ a_{0x}(\varphi, t) \varphi' (t) - b_{0x} (\varphi, t)
- c_{0 x} (\varphi, t) u_{0} (\varphi, t) - \frac{A(\varphi, t)}{c_0
(\varphi, t)} \right] = 0.
\end{equation}

\textbf{3.3. Differential equation for  discontinuity curve.} Using
(\ref{ort_cond}) as $ j =1 $ and (\ref{function_F}) by means of
tedious but not complicated calculations  we find the second order
ordinary differential equation for function $ \varphi = \varphi(t) $
in the following form
\begin{equation}\label{eq_rozryv}
\left[ A_1 \varphi{\,'\,^2} + A_2  \varphi{\,'} + A_3 \right]
\varphi{\,''} + A_4  \, \varphi{\,' \,^4} + A_5 \, \varphi{\,' \,
^3} + A_6 \, \varphi{\,' \,^2} + A_7 \, \varphi{\,'} = 0,
\end{equation}
where coefficients $  A_k = A_k(\varphi, t) $, $ k = \overline{1, 7}
$, are given as follows
$$
A_1 = 24 \, a_0^2 \, c_0, \, \, A_2 = - 8 \, a_0 \, c_0 \, \alpha ,
\, \, A_3  = - \, c_0 \, \alpha^2, \, \, A_4  = - 40 \, c_{0 x} \,
a_0^2 + 30 a_0 \, a_{0x} \, c_0 ,
$$
$$
A_5 = 60 \, a_0 \, c_{0x}\, \alpha + 20 \, a_0 \, a_{0t}\, c_0  - 24
\, a_{0}^2 \, c_{0 t}  - 30 \, a_0 \, c_0 \, \alpha_x - 15 \, a_{0x}
\, c_0 \, \alpha + 20 a_0 \, c_0^2 \, u_{0 x},
$$
$$
A_6 = - 20 \, a_0 \, c_0 \, \alpha_t \, - 5 \, a_{0t} \, c_0 \alpha
+ 15 \, c_0 \, \alpha \, \alpha_x  + 28 \, a_0 \, c_{0t} \, \alpha -
20 \, c_0^2 u_{0 x} \, \alpha - 20 \, c_{0x} \, \alpha^2,
$$
$$
A_7 = 5 \, c_0 \, \alpha \, \alpha_t - 20 \, c_{0t} \, \alpha^2,
$$
where $ \alpha = b_{0} + c_{0}  u_0 $, $ a_0 = a_0(\varphi, t) $, $
b_0 = b_0(\varphi, t)$, $ c_0 = c_0(\varphi, t) $, $ u_0 =
u_0(\varphi, t) $.

Differential equation (\ref{eq_rozryv}) is nonlinear and has smooth
coefficients. A problem of existence of its solution must be
studied in every case. In general, the equation % (\ref{eq_rozryv})
possesses a solution on only finite time interval. Therefore, the
supposed interval of existing its solution is denoted by $ [0;T]$.

\textbf{3.4. Exact solutions to system of differential equations
(\ref{sing_part_02}), (\ref{sing_part_j2}).} The general solution to
equation (\ref{sing_part_02}) in space $ G_1^0 $ is given by
formula (\ref{2sol_sing}). Let us proceed to searching a general
solution to inhomogeneous equation (\ref{sing_part_j2}) in exact
form through lemma 2 and lemma 3 providing us with necessary and
sufficient conditions of existing solutions to equation
(\ref{sing_part_j2}) in space $G_1$.

It is possible to find solution to (\ref{sing_part_j2}) after
getting a solution to equation (\ref{ad_eq_1}). The last one can be
solved using the method of variation of parameters because it's a
linear inhomogeneous ordinary differential equation. According to
described above procedure of constructing function $ v_0(t, \tau) $
in (\ref{2sol_sing}), function $ w_1 (t, \tau) = v_{0\tau} $ is a
non-trivial solution to homogeneous equation $ L w = 0 $. The other
linearly independent solution is given through Abel's formula
\cite{Teschl}
$$
w_2(t,\tau)= w_{1} (t, \tau) \int\limits_{\tau_0}^\tau w_{1}^{-2}(t,
\tau) \, d \tau = v_{0\tau} (t, \tau) \int\limits_{\tau_0}^\tau
v_{0\tau}^{-2}(t, \tau) \, d \tau .
$$

Thus, general solution to inhomogeneous equation
(\ref{sing_part_j2}) is written in the form
$$
v_j(t, \tau) = \left( \, \, \int\limits_{\tau_0}^\tau \Phi_j(t,
\tau_1) v_{0\tau} (t, \tau_1) \, d \tau_1 + C_3 \right) v_{0\tau}(t,
\tau) \int\limits_{\tau_0}^\tau v_{0\tau}^{-2} (t, \tau_1) \, d
\tau_1 -
$$
\begin{equation} \label{exact_v_j}
- \left( \, \, \int\limits_{\tau_0}^\tau \Phi_j(t, \tau_1) v_{0\tau}
(t, \tau_1)  \int\limits_{\tau_0}^{\tau_1} v_{0\tau}^{-2} (t, \xi)
\, d \xi \, d \tau_1 + C_4 \right) v_{0\tau}(t, \tau),
\end{equation}
where values $ C_3 $, $ C_4 $ are some real constants.

\textbf{3.5. Constructing terms of the singular part of asymptotic.}
At last functions $ V_j(x, t, \tau) $, $ j=\overline{0,N} $, can be
determined outside of the discontinuity curve $ \Gamma $. Taking
into consideration formulae (\ref{2sol_sing}), (\ref{exact_v_j})
providing us with their values on the curve $ \Gamma $ we define
functions $ V_j(x, t, \tau) $, $ j=\overline{0,N} $, through
extension $ v_j(t, \tau)$, $ j=\overline{0,N}$, from the curve $
\Gamma $ to its neighborhood.

Since $ v_0(t, \tau) \in G_1^0 $ we can put
\begin{equation} \label{prolong_V_0}
V_0(x, t, \tau) = v_0(t, \tau) .
\end{equation}

While extending $ V_j(x, t, \tau) $, $ j=\overline{1,N} $, two cases
should be considered. Firstly, we suppose condition
(\ref{ort_cond_2}) takes place, i.e. $ v_j(t, \tau)\in G_1^0 $. The
case is similar to the one for $ v_0(t, \tau) $. It means that
extension of function $ v_j(t, \tau)$, $ j = \overline{1, N} $, from
$ \Gamma $ to its neighborhood can be written as
\begin{equation} \label{prolong_V_j_0}
V_j(x, t, \tau) = v_j(t, \tau).
\end{equation}

In the opposite case when there isn't satisfied condition
(\ref{ort_cond_2}) we make use of representation (\ref{vyglyad}).
Thus extension of the function is realized as follows
\begin{equation} \label{prolong_V_j}
V_j(x, t, \tau) = u_j^- (x, t) \eta(t, \tau) + \psi_j(t, \tau),
\end{equation}
where functions $ \eta_j(t, \tau) $, $ \psi_j(t, \tau) $, $ j =
\overline{1, N} $, are defined under describing formulae
(\ref{vyglyad}), (\ref{nu}) while function $ u_j^- (x, t) $, $ j =
\overline{1, N} $, satisfies differential equation
\begin{equation}\label{prolong}
\Lambda u_j^- (x, t) = f_j^- (x, t), \quad j = \overline{1, N},
\end{equation}
\begin{equation}\label{operator_Lambda}
\Lambda = a_0(x, t) \frac{\partial}{\partial t} + b_0(x, t)
\frac{\partial}{\partial x} + c_0(x, t) u_0(x,
t)\frac{\partial}{\partial x} + c_0(x, t) u_{0 x}(x, t).
\end{equation}
In particular, here
$$
f_1^-(x, t) = 0, \quad f_2^-(x, t) = - a_1(x, t) \frac{\partial
u_1^-}{\partial t} - b_1(x, t) \frac{\partial u_1^-}{\partial x} -
c_1(x, t) u_1^- \frac{\partial u_0}{\partial x} -
$$
\begin{equation}\label{function_f}
- c_0(x, t) u_1^-\frac{\partial u_1}{\partial x} - c_0(x, t) u_1
\frac{\partial u_1^-}{\partial x} - c_0(x, t) u_1^- \frac{\partial
u_1^-}{\partial x} - c_1(x, t) u_0 \frac{\partial u_1^-}{\partial
x}.
\end{equation}

In addition, function $ u_j^-(x, t) $, $ j = \overline{1, N} $, is
clear to satisfy condition
\begin{equation}\label{prolong_0}
u_j^- (x, t)\biggr|_{\Gamma} = \nu_j(t), \quad j = \overline{1, N},
\end{equation}
following from (\ref{vyglyad}).

Thus, $ u_j^-(x, t) $, $ j = \overline{1, N} $, is a solution to the
Cauchy problem (\ref{prolong}), (\ref{prolong_0}).

In general, the Cauchy problem is well posed because the curve $
\Gamma $ is transversal to characteristics of the operator $ \Lambda
$. It follows that the problem has a solution in some neighborhood $
\Omega_{\varepsilon}(\Gamma) $ of the curve $ \Gamma $.

Therefore, the problem of constructing the singular part of asymptotic
(\ref{sol_one-phase}) is completely solved.

Summarizing the results given above, the following should be
noted. We have found the form of asymptotic solutions to the
singularly perturbed BBM equation with variable coefficients and
have described in details the algorithm for constructing such
solutions. Thus, we solved the first main problem of asymptotic
analysis, methods of which were successfully applied to obtain
approximate solutions of special form to the BBM equation
(\ref{BBM_v}).

\textbf{Remark 1.} {The constructed asymptotic soliton-like solution
to equation (1) doesn't belong to the Schwartz space in the general
case. It is a sum of regular part $ U_N(x,t,\varepsilon )$ and
singular part $ V_N(x,t,\tau , \varepsilon )$. The regular part $
U_N(x,t,\varepsilon )$ is only enough smooth function and it doesn't
belong to the Schwartz space in general.

If $ U_N(x,t,\varepsilon ) \equiv 0$ then the constructed asymptotic
solution may belong to the Schwartz space only in particular case,
because all terms of the singular part except the main term belong
to the space $ G_1 \not \subset G_0 $. The function $ V_N(x,t,\tau ,
\varepsilon )$ belongs to the Schwartz space only under the
condition (32) of lemma 3 in our paper. In general, the constructed
asymptotic solutions don't approximate the quickly decreasing
functions.}

\textbf{Remark 2.} {Through the proposed technique we can find
asymptotic solutions coinciding with exact soliton solutions in the
case of constant coefficients as well as we can construct the other
type of asymptotic solutions among them there are asymptotic
step-like solutions [51] and asymptotic $\Sigma$-solutions [52,53].
So, the set of constructed asymptotic solutions is wider than the
set of quickly decreasing solutions to the BBM equation
(\ref{BBM_v}).}

At present we turn to another important problem of asymptotic
analysis relating to justification of the constructed asymptotic
solutions.

\textbf{4. Precision of asymptotic solution (\ref{sol_one-phase}).}
Discussing the problem of asymptotic estimates for approximate
solutions built for equation (\ref{BBM_v}), we need to analyze the
accuracy with which the solutions satisfy this equation.

The asymptotic solutions for equation (\ref{BBM_v}) are represented
by formula (\ref{sol_one-phase}) where their singular part is
written in two ways depending on condition (\ref{ort_cond_2}).
However, both forms of the asymptotic solutions satisfy the equation
with the same precision as confirmed by the following relevant
statements.

\textbf{Theorem 1.} {\it Let the following conditions be supposed: \\
1. functions $ a_k(x, t) $, $ b_k(x, t) $, $ c_k (x, t) \in
C^{\infty}
({\bf R}\times [0; T])$, $ k = \overline{0, N} $; \\
2. inequality (\ref{A>0}) is fulfilled; \\
3. orthogonality conditions (\ref{ort_cond}) are satisfied; \\
4. conditions (\ref{ort_cond_2}) are realized.

Then asymptotic one-phase soliton-like solution to equation
(\ref{BBM_v}) is written as
\begin{equation}\label{as_sol_parn_11}
u_N (x, t, \varepsilon) = \left\{
\begin{array}{cl}
  Y_N^- (x, t, \varepsilon), & (x, t) \in D^- \backslash
  \Omega_{\varepsilon} (\Gamma), \\
  Y_N(x, t, \varepsilon), & (x, t) \in \Omega_{\varepsilon} (\Gamma), \\
  Y_N^+ (x, t, \varepsilon), & (x, t) \in D^+ \backslash \Omega_{\varepsilon} (\Gamma),
\end{array} \right.
\end{equation}
where
\begin{equation}\label{as_solY-}
Y_N^- (x, t, \varepsilon) = \sum\limits_{j=0}^N \varepsilon^j u_j(x,
t), \quad (x, t) \in D^- =\{(x, t)\in {\mathbf R} \times [0; T]: x -
\varphi(t) < 0\},
\end{equation}
\begin{equation}\label{as_solY}
Y_N (x, t, \varepsilon) = \sum\limits_{j=0}^N \varepsilon^j \left[
u_j(x, t) + V_j (t, \tau)\right], \quad \displaystyle \tau =
\frac{x-\varphi(t)}{\varepsilon}, \quad (x, t) \in
\Omega_{\varepsilon} (\Gamma),
\end{equation}
\begin{equation}\label{as_solY+}
Y_N^+ (x, t, \varepsilon) = \sum\limits_{j=0}^N \varepsilon^j u_j(x,
t), \quad (x, t) \in D^+ =\{(x, t)\in {\mathbf R} \times [0; T]: x -
\varphi(t) > 0\}.
\end{equation}

In addition, function (\ref{as_sol_parn_11}) satisfies equation
(\ref{BBM_v}) on the set $ {\mathbf R}\times [0; T] $ with accuracy
$ O(\varepsilon^N) $. As $ \tau \to \pm\infty $ the function
satisfies (\ref{BBM_v}) with asymptotic estimate $
O(\varepsilon^{N+1}) $.}

{\bf Proof of theorem 1.} Proving the theorem is not complicated and
it is done in the standard way similarly to the proof of the analogous
statement for the singularly perturbed KdV equation with variable
coefficients that was described in details in \cite{Sam3}. That is
why we avoid demonstration of tedious calculations in full here and
describe only the basic idea.

We have to obtain an asymptotic estimate for discrepancy of the
approximate solution that is given by formula
(\ref{as_sol_parn_11}).

From the structure of function (\ref{as_sol_parn_11}), we can see
that in domains $ D^+\backslash \Omega_{\varepsilon}(\Gamma) $, $
D^-\backslash \Omega_{\varepsilon}(\Gamma) $ asymptotic solution
contains only regular part of the asymptotic and in them it
satisfies the equation with accuracy $ O(\varepsilon^{N+1}) $
through its determining.

Thus, it remains only to consider domain $
\Omega_{\varepsilon}(\Gamma) $. Function (\ref{as_sol_parn_11})
is substituted into equation (\ref{BBM_v}). Further, equations
(\ref{reg_part_0}), (\ref{reg_part_1}) for regular part of
asymptotic (\ref{as_sol_parn_11}) are taken into account.

The next step is considering functions $ a_j(x, t), $ $ b_j(x, t) $,
$ c_j(x, t) $, $ u_j(x, t) $, $ j = \overline{0, N} $, in
neighborhood of the discontinuity curve $ \Gamma $ and their
representation as Taylor polynomials with the required accuracy.

Finally, equations (\ref{sing_part_02}), (\ref{sing_part_j2}) for
singular parts of asymptotic (\ref{as_sol_parn_11}) as well as
property of functions $ V_j(t, \tau) \in G_1^0 $, $ j = \overline{0,
N} $, are used for obtaining asymptotic estimation of discrepancy
for the approximate solution (\ref{as_sol_parn_11}). This completes
the proof of theorem 1.

{\bf Theorem 2.} {\it Let the following conditions be satisfied: \\
1. functions $ a_k(x, t) $, $ b_k(x, t) $, $ c_k (x, t) \in
C^{\infty}({\bf R}\times [0; T])$, $ k = \overline{0, N} $; \\
2. inequality (\ref{A>0}) is fulfilled; \\
3. orthogonality conditions (\ref{ort_cond}) are satisfied; \\
4. the Cauchy problem (\ref{prolong}), (\ref{prolong_0}) has a
solution on the set $ \{(x, t) \in{\mathbf R} \times[0; T]: x -
\varphi(t) \le 0 \}$.

Then the asymptotic one-phase soliton-like solution to equation
(\ref{BBM_v}) can be written as
\begin{equation}\label{as_sol_parn}
u_N (x, t, \varepsilon) = \left\{
\begin{array}{cl}
  Y_N^- (x, t, \varepsilon), & (x, t) \in D^- \backslash \Omega_{\varepsilon} (\Gamma), \\
  Y_N (x, t, \varepsilon), & (x, t) \in \Omega_{\varepsilon} (\Gamma), \\
  Y_N^+ (x, t, \varepsilon), & (x, t) \in D^+ \backslash \Omega_{\varepsilon} (\Gamma),
\end{array} \right.
\end{equation}
where
\begin{equation}\label{as_sol_parnY_-}
Y_N^- (x,t, \varepsilon) = \sum\limits_{j=0}^N \varepsilon^j
u_j(x,t) + \sum\limits_{j=1}^N \varepsilon^j u_j^- (x, t), \quad (x,
t) \in D^-,
\end{equation}
\begin{equation}\label{as_sol_parnY}
Y_N (x, t, \varepsilon) = \sum\limits_{j=0}^N \varepsilon^j \left[
u_j(x, t) + V_j (x, t, \tau)\right], \quad \displaystyle \tau =
\frac{x-\varphi(t)}{\varepsilon}, \quad (x, t) \in
\Omega_{\varepsilon} (\Gamma),
\end{equation}
\begin{equation}\label{as_sol_parnY_+}
Y_N^+ (x, t, \varepsilon) = \sum\limits_{j=0}^N \varepsilon^j u_j(x,
t), \quad (x, t) \in D^+.
\end{equation}

In addition, function (\ref{as_sol_parn}) satisfies equation
(\ref{BBM_v}) with accuracy $ O(\varepsilon^N) $ on the set $
{\mathbf R}\times [0; T] $. As $ \tau \to \pm\infty $ the solution
satisfies (\ref{BBM_v}) with asymptotic estimate $
O(\varepsilon^{N+1}) $.}

{\bf Proof of theorem 2.} Function (\ref{as_sol_parn}) gives another
form of asymptotic one-phase soliton-like solution to equation
(\ref{BBM_v}). However, the proof is similar to proof of theorem 1.
In other words, we have also to obtain asymptotic estimate for
discrepancy of the approximate solution that is given by formula
(\ref{as_sol_parn}).

In domains $ \Omega_\varepsilon(\Gamma) $, $ D^+ \backslash
\Omega_\varepsilon(\Gamma) $ function (\ref{as_sol_parn}) is the
same as (\ref{as_sol_parn_11}) through composition, in fact. So, the
proof of theorem 2 is analogous to proof of theorem 1 for the case
of these domains.

Finally, we consider set $ D^- \backslash \Omega_\varepsilon(\Gamma)
$. Here function (\ref{as_sol_parn}) differs from
(\ref{as_sol_parn_11}) in expression $ \sum\limits_{j=1}^N
\varepsilon^j u_j^- (x, t) $, $j=\overline{1,N} $. The last equates
relation (\ref{BBM_v}) with asymptotic precise $
O(\varepsilon^{N+1}) $, $ \varepsilon \to 0$.

Thus, on the set $ {\mathbf R}\times [0; T] $ function
(\ref{as_sol_parn}) satisfies equation (\ref{BBM_v}) with accuracy $
O(\varepsilon^N) $. Due to properties of functions forming the
singular part of the asymptotic solution, it satisfies equation
(\ref{BBM_v}) with precision $ O(\varepsilon^{N+1}) $ as $ \tau \to
\pm \infty $.

Theorem 2 is proved.

\textbf{5. Discussions and conclusions.}

This paper deals with one of the main problems of asymptotic analysis
that concerns the development of an algorithm for constructing
approximate (asymptotic) solutions to a partial differential
equation with singular perturbation. The solutions satisfy the
perturbed equation with a certain (asymptotic) accuracy.

We have found the form of the asymptotic solutions to the equation and
proposed and described in details an algorithm for constructing
asymptotic solutions of a special form for the singularly perturbed
Benjamin–Bona–Mahony (BBM) equation with variable coefficients.
Statements on justification of the algorithm are proved.

The asymptotic solutions consist of regular and singular parts.  The
singular part gives a solution that describes the soliton wave of
the BBM equation in the case of constant coefficients \cite{Wazwaz}.
In the case of variable coefficients for the BBM equation, a
singular part does not necessarily belong to a space of functions
that are quickly decreasing with respect to a phase variable.  But
in some cases, such a property is fulfilled. Therefore, similar
solutions can be considered as a generalization of soliton ones and
they can be called soliton-like \cite{Maslov1}.  The regular part
creates a background on which soliton-like waves move.

It should be also mentioned that the properties of solutions of
initial problems for evolution equations depend essentially on the
initial functions, especially in the case of nonlinear partial
differential equations. In this connection, we can mention the
Cauchy problem for the Korteweg-de Vries equation with smooth
initial function \cite{Pokhozhayev}. It has singular solutions that
are destroyed at finite time "roughly"\, or accordingly to scenario
of gradient catastrophe. The problem of existence of different type
of solutions to the Cauchy problem for the BBM equation is also
interesting, but it is not discussed in the paper. This challenging  problem
requires additional in-depth research.

The proposed algorithm can be applied to constructing asymptotic
soliton-like solutions of different nonlinear partial differential
equations of integrable type with variable coefficients because the
algorithm allows us to obtain exact soliton solutions to the equations
in the case of constant coefficients.

Moreover, the proposed technique allows us to construct not only
asymptotic soliton-like solutions coinciding with exact soliton
solutions in case of constant coefficients but also the other types
of approximate solutions, including asymptotic step-like
\cite{Khruslov} and asymptotic $\Sigma$ -- solutions \cite{Sam5},
\cite{Sam6}.

{\bf Acknowledgements.} This research was partially supported by
Ministry of Education and Science of Ukraine and Taras Shevchenko
National University of Kyiv [grant number 18~BA~038~--~01].

Authors would like to convey their cordial thanks to Prof. Roman
Samulyak (Stony Brook University, New York, USA) for the discussion
of the paper and useful suggestions.

Authors are also thankful to anonymous Referees who have attentively
read the manuscript and made very important remarks and comments.

\renewcommand{\refname}{References}

\hfill Revised 23 November 2018

\end{document}